\documentclass[11pt]{article}
\usepackage{acl2014}
\usepackage{times}
\makeatletter
\newcommand{\@BIBLABEL}{\@emptybiblabel}
\newcommand{\@emptybiblabel}[1]{}
\makeatother
\PassOptionsToPackage{hyphens}{url}\usepackage[pdftex]{hyperref}
\usepackage{latexsym}
\usepackage{comment}
\usepackage[pass,paperwidth=21cm,paperheight=29.7cm]{geometry}
\usepackage{color}
\usepackage{graphicx}
\usepackage{subfigure}
\usepackage{xspace}
\usepackage{caption}
\usepackage[clock]{ifsym} %
\usepackage{mathabx} %
\usepackage{multirow} %
\usepackage[small,compact]{titlesec} %
\titlespacing{\subsection}{0pt}{2pt}{2pt}
\titlespacing{\paragraph}{0pt}{2.3pt}{0pt}
\let\oldpara\paragraph
\renewcommand{\paragraph}[1]{\oldpara{#1}~}

\newif{\ifhidecomments}
\hidecommentsfalse   %

\def\custom{custom\xspace}

\title{
The effect of wording on message propagation: \\
Topic- and author-controlled natural experiments on Twitter
}

\author{Chenhao Tan\\
  Dept. of Computer Science \\
  Cornell University \\
  {\href{mailto:chenhao@cs.cornell.edu}{chenhao@cs.cornell.edu}} \\\And
  Lillian Lee \\
  Dept. of Computer Science \\
  Cornell University \\
  {\href{mailto:llee@cs.cornell.edu}{llee@cs.cornell.edu}} \\ \And
  Bo Pang \\
  Google Inc. \\
  {\href{bopang42@gmail.com}{bopang42@gmail.com}} \\}

\date{}

\begin{document}
\maketitle

\begin{abstract}
Consider a person trying to spread an important message on a social network.
He/she can spend hours trying to craft the message.
Does it actually matter?
While there has been extensive prior work looking into predicting popularity of social-media content,
the effect of wording per se has rarely been studied
since it is often confounded
with
the popularity of the author and the topic.
To control for these confounding factors,
we take advantage of the surprising fact that there are
many
pairs of tweets containing the
{\em same} url and written by the {\em same} user but employing different
wording.
Given such pairs, we ask:
which version attracts more retweets?
This turns out to be a more difficult task than predicting popular topics.
Still, humans can answer this question better than chance (but far from
perfectly), and the computational methods we
develop
can do better than both an average human
and a strong competing method
trained on non-controlled data%
.

\end{abstract}

\section{Introduction}
\label{sec:intro}
How does one make a message ``successful''?
This question is of interest to
many entities, including political parties trying to {\em frame} an
issue \cite{Chong+Druckman:2007a}, and individuals attempting to make
a point in a group meeting.  In the first case, an important type of
success is achieved if the national conversation adopts the rhetoric
of the party; in the latter case, if other group members repeat the
originating individual's point.

The massive availability of online
messages, such as posts to social media, now affords researchers new means to
investigate at a very large scale the factors affecting message
propagation,
also known as adoption, sharing,  spread, or virality.
According to prior research, important features include characteristics of the originating author (e.g., verified
Twitter
user or not, author's messages' past success rate), the author's
social network
(e.g., number of followers),  message timing, and message
content or topic
\cite{artzipredicting,Bakshy:ProceedingsOfWsdm:2011,Borghol+etal:12,Guerini:ProceedingsOfIcwsm:2011,Guerini:ProceedingsOfIcwsm:2012,hansen2011good,Hong:2011:PPM:1963192.1963222,Lakkaraju+McAuley+Leskovec:13,Berger+Milkman:12,ma2012will,petrovic2011rt,Romero+Tan+Ugander:13,Suh+etal:10,Sun+Zhang+Mei:13,Tsur+Rappoport:12}.
Indeed,
it's not
surprising that
one of the most retweeted tweets of all time
was
from
user BarackObama, with 40M followers,  on November 6, 2012:
``Four more years. [link to photo]''.

Our interest in this paper is the effect of
alternative
message {\em wording}, meaning {\em how} the message is said, rather than what the
message is about.
In contrast to the identity/social/timing/topic features
mentioned above,  wording is
one of the few factors
directly under an author's control
when he or she seeks to convey a
{\bf fixed} piece of content.  For example, consider a speaker at the ACL business
meeting who has been tasked with  proposing that Paris be the next ACL location.  This person cannot on the
spot become ACL president,  change the shape of
his/her social network, wait until the next morning to
speak, or campaign for Rome instead;
but
he/she can craft
the
 message to be
more humorous, more informative,
emphasize certain aspects instead of others,
and so on.
In other words, we investigate whether a different choice of words
affects message propagation, \emph{controlling for user and
topic}: would user BarackObama have gotten significantly more (or
fewer) retweets if he had used some alternate wording to announce his
re-election?
\newcommand{\tuc}{\mbox{TAC}\xspace}

\begin{table*}[htb!]
\centering
\caption{
Topic- and author-controlled (\tuc) pairs. Topic control = inclusion
of the same URL.
\label{tb:tweet_example}}
\scriptsize
\begin{tabular}{|l|p{13cm}|l|}
\hline
 author %
 & tweets &\#retweets \\ \hline
natlsecuritycnn %
& $t_1$: FIRST ON CNN: After Petraeus scandal, Paula Broadwell looks to recapture `normal life.' {\bf http://t.co/qy7GGuYW} & $n_1$ = 5 \\
\cline{2-3} &
$t_2$: First on CNN: Broadwell photos shared with Security Clearance as she and her family fight media portrayal of her {\bf [same URL]} & $n_2$ = 29\\
\hline \hline
ABC
& $t_1$: Workers, families take stand against Thanksgiving hours: {\bf http://t.co/J9mQHiIEqv} & $n_1$ = 46\\
\cline{2-3}& $t_2$: Staples, Medieval Times Workers Say Opening Thanksgiving Day Crosses the Line {\bf [same URL]}  &  $n_2$ = 27\\
\hline \hline
cactus\_music
& $t_1$: I know at some point you've have been saved from hunger by our rolling food trucks friends. Let's help support them! {\bf http://t.co/zg9jwA5j} & $n_1$ = 2\\
\cline{2-3} & $t_2$: Food trucks are the epitome of small independently owned LOCAL businesses! Help keep them going! Sign the petition {\bf [same URL]} &  $n_2$ = 13\\

\hline
\end{tabular}
\end{table*}

Although we cannot create a parallel universe in which BarackObama
tweeted something else\footnote{%
Cf.\ the Music Lab ``multiple
universes'' experiment to test the randomness of popularity
\cite{Salganik:Science:2006}.},
fortunately, a surprising characteristic of Twitter allows us to run a
fairly analogous {\em natural experiment}:  external forces
serendipitously provide an environment that resembles the desired
controlled setting \cite{dinardo2008natural}.  Specifically, {\em it turns out to be
unexpectedly common for the same user to post different tweets regarding the
same URL} --- a good proxy for fine-grained topic\footnote{%
Although hashtags have been used as coarse-grained topic labels in prior work,
for our purposes, we have no assurance that two tweets both using, say,
``\#Tahrir'' would be attempting to express the same message but in different
words.  In contrast, see the same-URL examples in Table
\ref{tb:tweet_example}.}
--- within a relatively short period of time.\footnote{Moreover, Twitter presents tweets to a reader in
strict chronological order, so that there are no algorithmic-ranking effects to
compensate for in determining whether readers saw a tweet.  And, Twitter accumulates retweet counts for the entire retweet
cascade and displays them for the original tweet at the root of the propagation
tree, so we can directly use Twitter's retweet counts to compare the entire reach of
the different versions.}
Some example pairs are shown in
Table \ref{tb:tweet_example}; we see that the paired tweets may differ
dramatically, going far beyond word-for-word substitutions, so that quite
interesting changes can be studied.

Looking at these examples,
can one in fact tell from the
wording which
tweet in
a topic- and author-controlled pair will be more successful?  The answer
may not be a priori clear.
For example,  for the first pair in the table,
one person we asked found
$t_1$'s
invocation of a ``scandal'' to be
more attention-grabbing;
but another person preferred
$t_2$
because it is more informative about the URL's content and includes ``fight
media portrayal''.  In an Amazon Mechanical Turk (AMT) experiment
(\S \ref{sec:turkers}), we found that humans achieved an average accuracy of
61.3\%: not that high, but better than chance, indicating that it is somewhat possible for humans
to predict greater message spread from different deliveries of the same
information.

Buoyed by the evidence of our AMT study that
wording effects exist, we then performed a battery of
experiments to seek generally-applicable, non-Twitter-specific features of more
successful phrasings.  \S \ref{sec:testing} applies hypothesis testing
(with Bonferroni correction
to ameliorate issues with multiple comparisons)
to  investigate the utility of features like
informativeness, resemblance to headlines,
and conformity to the community norm in language use%
.
\S \ref{sec:pred} further validates our findings via
prediction experiments, including on completely fresh held-out data, used only
once and after an array of
standard cross-validation
experiments.\footnote{And after crossing our fingers. %
}
We achieved
66.5\% cross-validation accuracy and 65.6\% held-out accuracy with a
combination of our custom features and bag-of-words.
Our classifier fared significantly better than
a number of baselines, including a
strong
classifier trained on the most-
and least-retweeted tweets that was
even granted
access to author and timing metadata.
\section{Related work}
\label{sec:related}

The idea of using carefully controlled experiments to study
effective communication
strategies
dates back
at least
to \newcite{Hovland+Janis+Kelley:54}.
Recent studies range from
examining what characteristics of \emph{New York Times} articles correlate with high re-sharing rates
 \cite{Berger+Milkman:12}
to
looking at how differences in description affect the spread of content-controlled
videos or images \cite{Borghol+etal:12,Lakkaraju+McAuley+Leskovec:13}.
\newcite{simmons2011memes} examined the variation of quotes from
different sources
to
examine how
textual memes mutate as people pass them along,
but did not control for
author.
Predicting the ``success'' of various texts such as novels
and movie quotes has been the aim of additional prior work not already mentioned in \S\ref{sec:intro}
\cite{Ashok+Feng+Choi:13,Louis+Nenkova:13,Danescu-Niculescu-Mizil+Cheng+Kleinberg+Lee:12,Pitler:2008,McIntyre:2009:LTT:1687878.1687910}.
To
our knowledge,
there have been no large-scale studies
exploring wording effects in
a
both topic- and author-controlled setting.
Employing such controls, we find
that
 predicting the more effective
alternative wording is much harder than the previously well-studied
problem of predicting popular content
when author or topic can freely vary.

Related work
regarding the
features
we considered
is
deferred to
\S\ref{sec:testing} (features description).
\newcommand{\condexp}[2]{\widehat{E}(#1|#2)}

\section{Data}
\label{sec:data}

\newcommand{\better}{better\xspace}
\newcommand{\worse}{worse\xspace}
Our main dataset was constructed by first gathering 1.77M
topic- and author-controlled (henceforth {\em \tuc}) tweet pairs\footnote{%
No data collection/processing was conducted at Google.}
differing in more than just spacing.\footnote{The total excludes: tweets containing multiple
URLs; tweets from users posting about the same URL more than five times (since
such users might be spammers); the third, fourth, or fifth version for users posting
between three and five
tweets for the same URL; retweets (as identified by Twitter's API or by
beginning with ``RT @''); non-English tweets.}
We accomplished this by crawling
timelines of 236K user ids
that appear
 in
prior work \cite{Kwak+Lee+Park+Moon:10,Yang:2011:PTV:1935826.1935863}
via the Twitter API.
This crawling process also yielded 632K \tuc pairs whose
only difference was
spacing, and an additional
558M ``unpaired'' tweets;
as shown later in this paper, we
used these extra corpora for computing language models and other auxiliary information.
We applied
non-obvious but important filtering --- described later in
this section ---  to control for other external factors
and
to reduce ambiguous cases.
This brought us to a set of
11,404 pairs,
with the {\em gold-standard} labels determined by which tweet in each pair was the one that received
more retweets according to the Twitter API.
We then did a second crawl to get an additional
1,770 pairs to serve as a held-out dataset.
The corresponding tweet IDs are available online at
\url{http://chenhaot.com/pages/wording-for-propagation.html}.
(Twitter's terms of service prohibit sharing the actual tweets.)
\begin{figure}[t]
\centering
\subfigure[For {\em identical} \tuc pairs, retweet-count deviation vs. time lag between $t_1$ and $t_2$, for the author follower-counts given in the legend.] {
  \includegraphics[width=0.22\textwidth]{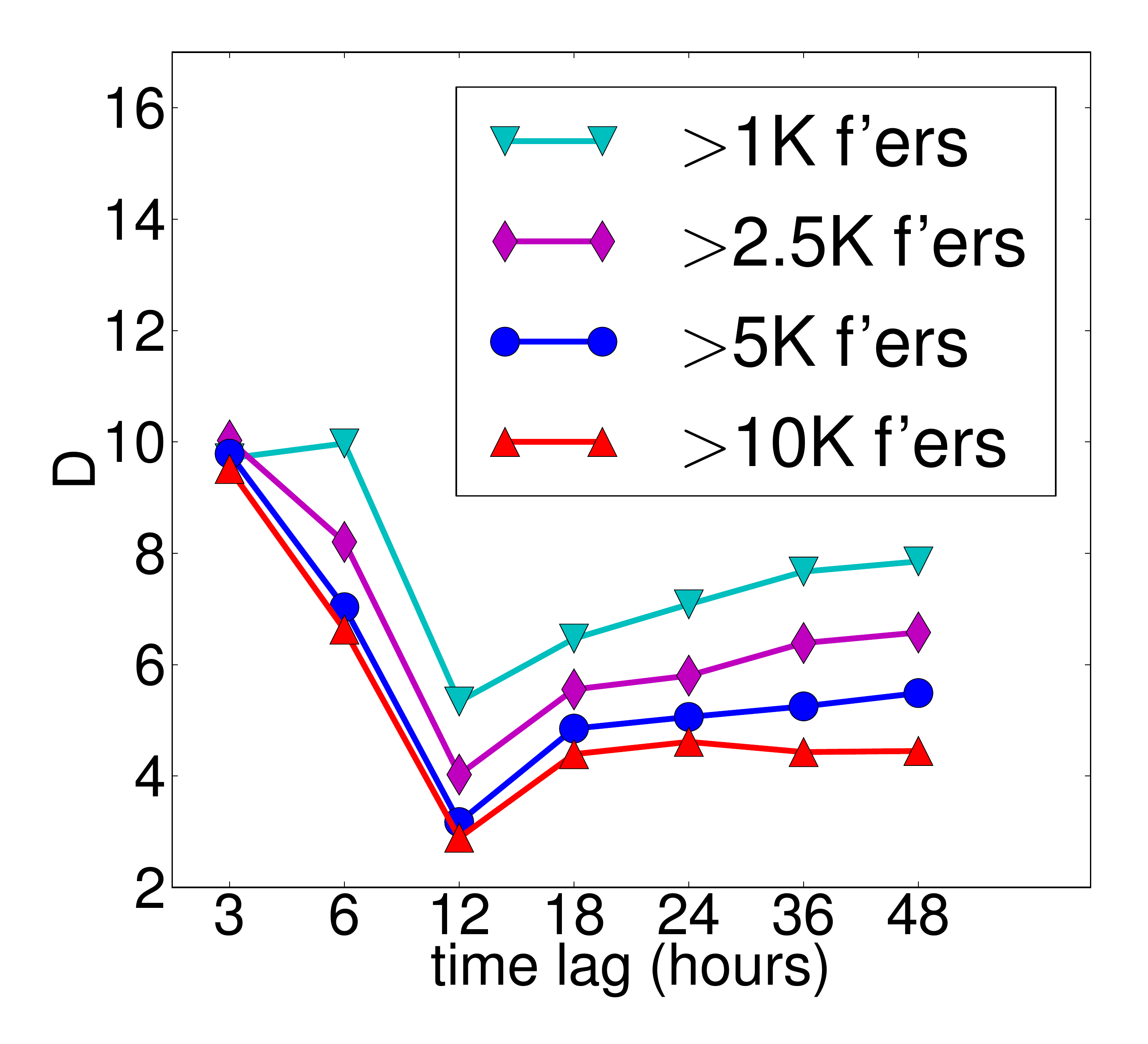}
  \label{fig:difference_varied}
}%
\hfill
\subfigure[
Avg. $n_2$ vs.\ $n_1$ for identical \tuc pairs, highlighting our chosen time-lag and
follower thresholds.
Bars: standard error.
Diagonal line: $\condexp{n_2}{n_1}=n_1$.
] {
  \includegraphics[width=0.22\textwidth]{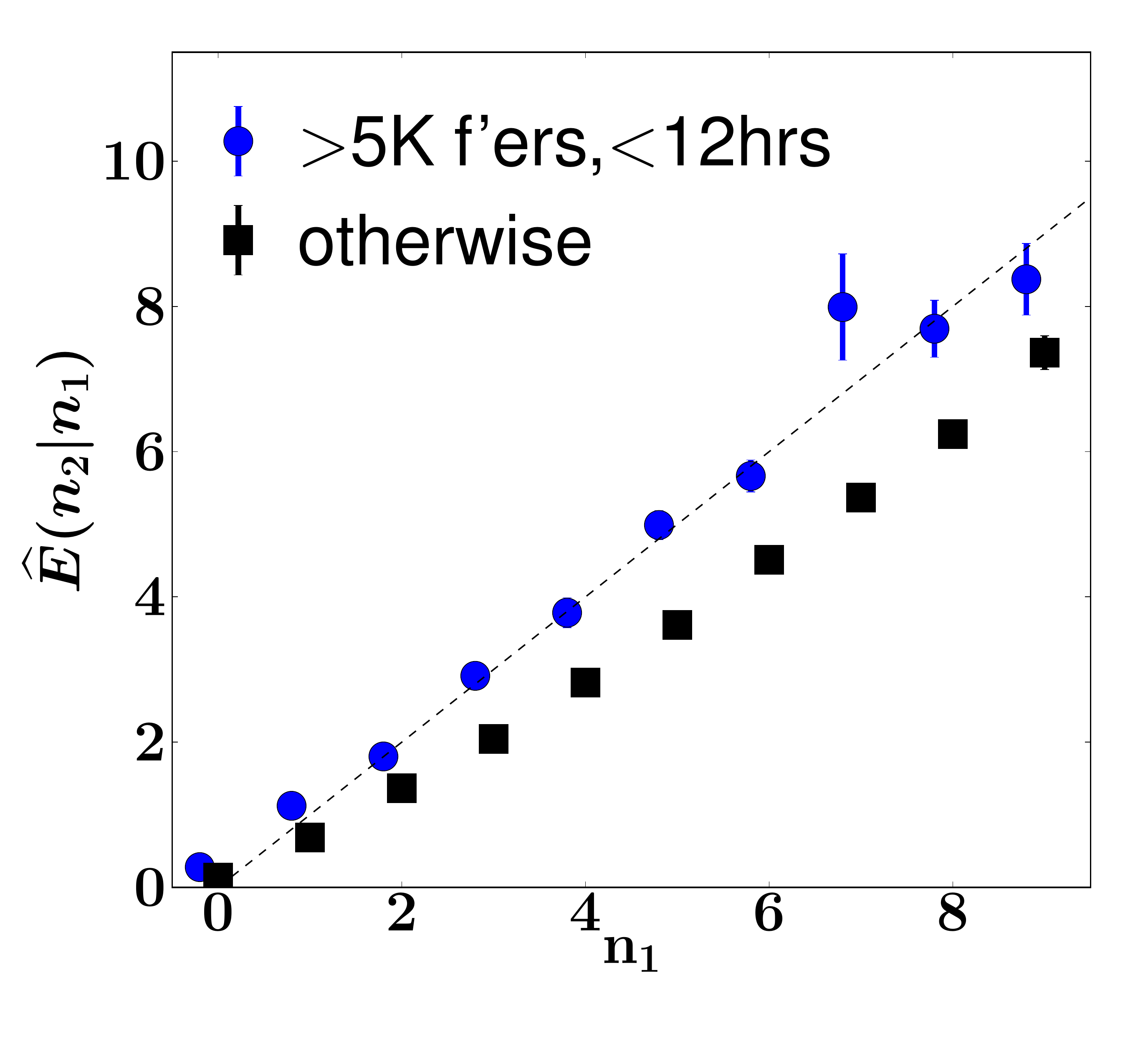}
  \label{fig:difference_same_n1}
}
\caption{
(a): The ideal case where $n_2 = n_1$ when $t_1 = t_2$ is best approximated when
$t_2$
occurs within 12 hours of $t_1$ and the author has at least 10,000 or 5,000 followers.
(b): in our chosen setting (blue circles), $n_2$ indeed tends to track
$n_1$, whereas otherwise (black squares), there's a bias towards
retweeting $t_1$.
\label{fig:same_n1_n2}}
\end{figure}

Throughout, we refer to the
textual content of the
earlier tweet within a
\tuc
pair as
$t_1$, and
of
the later
one as $t_2$.
We denote the number of retweets received by each tweet  by $n_1$ and $n_2$,
respectively.
We refer to the tweet with higher (lower)
$n_i$ as the
``\better (\worse)'' tweet.
\paragraph{Using ``identical'' pairs to
determine how to
compensate for follower-count and
timing effects.}
In an ideal setting,
differences between $n_1$ and $n_2$
would be determined solely by
differences in wording.
But even with a \tuc pair, retweets might exhibit a temporal bias because of
the chronological order of tweet presentation ($t_1$ might enjoy a first-mover advantage \cite{Borghol+etal:12} because it is the ``original''; alternatively, $t_2$ might be
preferred because retweeters consider $t_1$ to be ``stale''). Also, the number of
followers an author has can
have complicated indirect effects on which
tweets are read (space limits preclude discussion).

We use the 632K \tuc pairs wherein $t_1$ and $t_2$ are {\em
identical}\footnote{Identical up to spacing: Twitter prevents exact copies by
the same author appearing within a short amount of time, but some authors work
around this by inserting spaces.} to check for  such confounding
effects: we see how much $n_2$ deviates from $n_1$ in such settings, since if
wording were the only explanatory factor, the retweet rates for identical
tweets ought to be equal.  Figure \ref{fig:difference_varied} plots how the
time
lag between $t_1$ and $t_2$ and the author's follower-count affect
the following deviation estimate:
$$D =\sum_{0 \leq n_1 < 10}|\condexp{n_2}{n_1}-n_1|,$$
\noindent where $\condexp{n_2}{n_1}$ is the average value of $n_2$ over pairs
whose $t_1$ is retweeted $n_1$ times.
(Note that the number of
pairs whose $t_1$ is retweeted $n_1$ times decays exponentially with $n_1$; hence, we condition on $n_1$
to keep the estimate from being dominated by pairs with
$n_1=0$, and do not consider $n_1 \geq 10$ because there are too few such pairs
to estimate
$\condexp{n_2}{n_1}$ %
reliably.)
Figure \ref{fig:difference_varied} shows %
that the setting where we (i) minimize the confounding effects
of time
lag
and
author's follower-count and (ii) maximize the amount of data to
work with is: when $t_2$
occurs
within 12 hours after $t_1$ and the author has
more than 5,000 followers.
Figure \ref{fig:difference_same_n1} confirms that for identical \tuc pairs,
our chosen setting indeed results in $n_2$ being on average close to
$n_1$, which corresponds to the
desired setting where wording is the dominant differentiating factor.%
\footnote{We also computed the Pearson correlation between $n_1$ and $n_2$,
even though it can be dominated by pairs with smaller $n_1$.
The correlation is 0.853
for %
``$>5K$ f'ers, $<$12hrs'', clearly higher than the 0.305 correlation for ``otherwise''.}

\paragraph{Focus on meaningful and general  changes.} Even after
follower-count and time-lapse filtering, we still
want to focus
on \tuc pairs
that
(i) exhibit
significant/interesting
textual changes (as exemplified in Table \ref{tb:tweet_example}, and as opposed to typo
corrections and the like), and (ii) have $n_2$ and
$n_1$ sufficiently different
so that we are confident in which $t_i$ is better
at attracting retweets.  To take care of (i), we discarded the
50\% of pairs
whose similarity was above the median, where similarity was tf-based cosine.\footnote{Idf weighting
was not employed because changes to frequent words are of potential interest. Urls, hashtags, @-mentions and numbers were normalized to [url], [hashtag], [at],
and [num] before computing similarity.}
For (ii), we sorted the remaining pairs by $n_2-n_1$ and retained only the top
and bottom 5\%.\footnote{For our data, this meant $n_2-n_1 \geq 10$ or
$\leq -15$. Cf. our median number of retweets: 30.}
Moreover, to ensure that we do not overfit to the idiosyncrasies of
particular authors, we cap the number of pairs contributed by each author to 50
before we deal with (ii).

\section{Human accuracy on \tuc pairs}
\label{sec:turkers}

We first ran a pilot study on Amazon Mechanical Turk (AMT) to determine
whether humans can identify, based on wording differences alone, which of two topic- and author- controlled tweets
is spread more widely. Each of our 5 AMT tasks
involved a disjoint set of 20 randomly-sampled \tuc pairs (with $t_1$ and
$t_2$ randomly reordered); subjects indicated ``which tweet would other
people be more likely to retweet?'', provided a short justification for their
binary response, and clicked a checkbox if they found that their choice was a
``close call''. We received 39
judgments
 per pair in aggregate from 106 subjects total
(9 people completed all 5 tasks).
The subjects' justifications were of very high quality, convincing us that they all did the task in good faith\footnote{We also note that the feedback we got was quite positive, including:  ``...It's fun to make choices between close tweets
  and use our subjective opinion.  Thanks and best of luck with your
  research'' and ``This was very interesting and really made me think
  about how I word my own tweets.  Great job on this survey!''. We
  only had to  exclude one person (not counted among the 106 subjects), doing so because he or she gave the same uninformative
  justification
  for all pairs.}.
Two examples for the third \tuc pair in Table \ref{tb:tweet_example} were:
 ``[$t_1$ makes] the cause relate-able
to some people, therefore showing more of an appeal as to why should
they click the link and support'' and, expressing the opposite view,  ``I like [$t_2$] more
because [$t_1$] starts out with a generalization that doesn't affect
me and try to make me look like I had that experience before''.
If we view the set of 3900 binary judgments for our 100-\tuc-pair sample as
constituting independent responses, then the accuracy for this set is 62.4\%
(rising to 63.8\% if we exclude the 587 judgments deemed ``close calls'').
However, if we evaluate the accuracy of the {\em majority} response among the
39
judgments per pair, the number rises to 73\%.
The accuracy of the majority response generally increases with the dominance
of the majority, going above 90\% when at least 80\% of the judgments agree
(although less than a third of the pairs satisfied this criterion).

Alternatively, we can consider the average accuracy of the 106 subjects: 61.3\%, which is better than chance
but far from 100\%.  (Variance was high:
one subject achieved 85\% accuracy out of 20 pairs, but eight scored below
50\%.) This result is noticeably lower than the 73.8\%-81.2\% reported by \newcite{petrovic2011rt}, who ran a similar experiment involving two subjects and 202 tweet pairs, but
where the pairs were {\em not} topic- or
author-controlled.\footnote{The accuracy range stems from whether
author's social features were supplied and which subject was considered.}

We conclude that even though propagation prediction becomes more
challenging when topic and author controls are applied, humans can
still to some degree tell which wording attracts more retweets.
Interested readers can try this out themselves at \url{http://chenhaot.com/retweetedmore/quiz}.

\newcommand{\omt}[1]{#1}
\newcommand{\bigomt}[1]{} %
\section{Experiments}
\label{sec:exp}
\newcommand{\rs}{rs}
\newcommand\shortrs{rt score\xspace}
\newcommand{\white}[1]{\textcolor{white}{#1}}

\newcommand{\eff}{effective?\xspace}  %
\newcommand{\perc}{$\%(f_2 > f_1)$\xspace}%
\newcommand{\su}{author-preferred?\xspace}%
\newcommand{\uses}{\#difference\xspace}

\omt{
  \begin{table*}[t]
  \centering
  \caption{Notational conventions for tables in
    \S\ref{sec:testing}.
   \label{tb:notation}
   }
    {%
  \begin{tabular}{cc}
    \begin{tabular}
      {l@{: }ll@{: }l}
      \multicolumn{4}{c}{{\it One-sided paired t-test for feature efficacy}} \\
       \hline
      $\uparrow\uparrow\uparrow\uparrow$ & p$<$1e-20 & $\downarrow\downarrow\downarrow\downarrow$ & p$>$1-1e-20 \\
      $\uparrow\uparrow\uparrow$ & p$<$0.001 & $\downarrow\downarrow\downarrow$ & p$>$0.999 \\
      $\uparrow\uparrow$ & p$<$0.01  & $\downarrow\downarrow$ & p$>$0.99 \\
      $\uparrow$ & p$<$0.05  & $\downarrow$ & p$>$0.95  \\
      \multicolumn{4}{l}{$*$: passes our Bonferroni correction}
    \end{tabular}
    &
    \begin{tabular}{l@{: }l}
    \multicolumn{2}{c}{{\it One-sided binomial test for feature increase}} \\
    \multicolumn{2}{c}{{\it (Do authors prefer to `raise' the feature in $t_2$?)}} \\ \hline
    YES & $t_2$ has a higher feature score than $t_1$,  $\alpha = .05$\\
    NO &  $t_2$ has a lower feature score than $t_1$, $\alpha = .05$\\
    (x\%) & \perc, if sig. larger or smaller  than 50\% \\
    \multicolumn{2}{c}{~} \\ %
    \end{tabular}
  \end{tabular}
  } %
  \end{table*}
} %

\bigomt{  %
\begin{table*}[t]
\caption{Testing on different groups of features.
 Testing for feature having higher value in the better tweet:
 $p<0.001:\uparrow\uparrow\uparrow$,
 $p<0.01:\uparrow\uparrow$,
 $p<0.05:\uparrow$,
 $p>0.999:\downarrow\downarrow\downarrow$,
 $p>0.99:\downarrow\downarrow$,
 $p>0.95:\downarrow$.
 Italics and $*$ indicates passing the Bonferroni correction.
 Testing for feature having higher value in the second tweet:
 we label ``YES'' (or ``NO'')
 respectively if this is true for more (or less) than 50\% of all tweets
 under confidence level 0.05.\label{tb:test}}
\small
\subtable[Explicit requests and Informativeness]{
  \centering
  \label{tb:test_call_action}
  \begin{tabular}{lll}
     \hline
rt & $\uparrow\uparrow\uparrow\uparrow\white{}$~* &------\\
retweet & $\uparrow\uparrow\uparrow\uparrow\white{}$~* &YES~(59\%)\\
spread & $\uparrow\uparrow\uparrow\white{\uparrow}$~* &YES~(56\%)\\
please & $\uparrow\uparrow\uparrow\white{\uparrow}$~* &------\\
pls & $\uparrow\white{\uparrow\uparrow\uparrow}$ &------\\
plz & $\uparrow\uparrow\white{\uparrow\uparrow}$ &------\\

     \hline\hline
length (chars) & $\uparrow\uparrow\uparrow\uparrow\white{}$~* &YES~(54\%)\\
verb & $\uparrow\uparrow\uparrow\uparrow\white{}$~* &YES~(56\%)\\
noun & $\uparrow\uparrow\uparrow\uparrow\white{}$~* &------\\
adjective & $\uparrow\uparrow\uparrow\white{\uparrow}$~* &YES~(51\%)\\
adverb & $\uparrow\uparrow\uparrow\white{\uparrow}$~* &YES~(55\%)\\
proper noun & $\uparrow\uparrow\uparrow\white{\uparrow}$~* &NO\white{--}~(45\%)\\
number & $\uparrow\uparrow\uparrow\uparrow\white{}$~* &NO\white{--}~(48\%)\\
hashtag & $\uparrow\white{\uparrow\uparrow\uparrow}$ &------\\
@-mention & $\downarrow\downarrow\downarrow\white{\downarrow}$~* &YES~(53\%)\\

     \hline
  \end{tabular}
}
\subtable[Provocativeness, sentiment, and language models
]{
  \centering
  \label{tb:test_retweet_score}
  \begin{tabular}{lll}
    \hline
\shortrs & $\uparrow\uparrow\white{\uparrow\uparrow}$~* &NO\white{--}~(49\%)\\
verb \shortrs & $\uparrow\uparrow\uparrow\uparrow\white{}$~* &------\\
noun \shortrs & $\uparrow\uparrow\uparrow\white{\uparrow}$~* &------\\
adjective \shortrs & $\uparrow\white{\uparrow\uparrow\uparrow}$ &YES~(50\%)\\
adverb \shortrs & $\uparrow\white{\uparrow\uparrow\uparrow}$ &YES~(51\%)\\
proper noun \shortrs & -------- &NO\white{--}~(48\%)\\

    \hline\hline
 positive & $\uparrow\uparrow\uparrow\white{\uparrow}$~* &------\\
 negative & $\uparrow\uparrow\uparrow\white{\uparrow}$~* &------\\
 contrast & $\uparrow\uparrow\uparrow\white{\uparrow}$~* &------\\

    \hline\hline
headline unigram & $\uparrow\uparrow\white{\uparrow\uparrow}$ &YES~(53\%)\\
headline bigram & $\uparrow\uparrow\uparrow\uparrow\white{}$~* &YES~(52\%)\\

twitter unigram & $\uparrow\uparrow\uparrow\white{\uparrow}$~* & YES~(54\%)\\
twitter bigram & $\uparrow\uparrow\uparrow\white{\uparrow}$~* & YES~(52\%)\\

personal unigram & $\uparrow\uparrow\uparrow\white{\uparrow}$~* &YES~(52\%)\\
personal bigram & -------- &NO\white{--}~(48\%)\\

  \hline
  \end{tabular}
}
\subtable[Pronouns, generality, and readability.
]{
  \begin{tabular}{lll}
1st person singular & -------- &YES~(51\%)\\
1st person plural & -------- &YES~(52\%)\\
2nd person & -------- &YES~(57\%)\\
3rd person singular & $\uparrow\uparrow\white{\uparrow\uparrow}$ &YES~(55\%)\\
3rd person plural & $\uparrow\white{\uparrow\uparrow\uparrow}$ &YES~(58\%)\\

\hline\hline
indefinite articles (a,an) & $\uparrow\uparrow\uparrow\white{\uparrow}$~* &------\\
definite articles (the) & -------- &YES~(52\%)\\

\hline\hline
reading ease & $\uparrow\uparrow\white{\uparrow\uparrow}$ &YES~(52\%)\\
negative grade level & $\uparrow\white{\uparrow\uparrow\uparrow}$ &YES~(52\%)\\

\hline\hline

  \end{tabular}
}
\end{table*}
} %

We now investigate computationally what wording features correspond to
messages achieving a broader reach.
We start (\S\ref{sec:testing}) by introducing a set of generally-applicable and (mostly) non-Twitter-specific features
to capture our intuitions
about what might be better ways
to phrase
a message.
We
then
use hypothesis testing (\S\ref{sec:testing}) to evaluate the importance of each feature for message propagation
and the extent to which authors employ it,
followed by experiments on a prediction task (\S\ref{sec:pred}) to
further examine the utility of these features.
\subsection{Features: efficacy and author preference}
\label{sec:testing}

\newcommand{\ttablesize}{\small}

What kind of phrasing
helps message propagation?
Does it work to explicitly ask people to share the message?
Is it better to be short and concise or long and informative?
We define an array of features to capture these
and other
messaging aspects.
We
then examine (i) how effective each feature is for
attracting more retweets; and (ii) whether authors prefer applying a
given
feature
when issuing a second version of a tweet.
\newcommand{\shortBC}{BC\xspace}
\newcommand{\metrics}{features\xspace}
\newcommand{\metric}{feature\xspace}
First, for each \metric, we use a one-sided paired t-test
to test whether, on our 11K \tuc pairs,  our score function for that \metric is larger in the better tweet versions than in the
worse tweet versions, for significance levels $\alpha = .05, .01,
.001
, 1\mbox{e-}20$.
Given that we did 39 tests in total, there is a risk of
obtaining false positives due to
multiple testing \cite{dunn1961multiple,benjamini1995controlling}.
To account for this, we also
report significance results for the conservatively Bonferroni-corrected (``\shortBC'')
significance level
$\alpha$ = 0.05/39=1.28e-3.
Second,
we examine author preference for applying a feature.
We do so because one (but by no means the only) reason authors post $t_2$
after having already advertised the same URL in $t_1$ is that these authors
were dissatisfied with the amount of attention
$t_1$ got; in such cases, the
changes may  have been specifically intended to attract more retweets.
We measure author preference for a \metric by the percentage of our \tuc
pairs\footnote{
For our preference experiments,
   we added in pairs where $n_2 - n_1$ was not in the top or bottom 5\% (cf.
   \S\ref{sec:data}, meaningful changes),
   since to measure author preference it's not necessary that the retweet
   counts differ significantly.
}
 where
$t_2$ has more ``occurrences'' of the feature than $t_1$,
which we denote by ``\perc''.
We use the one-sided binomial test to see whether \perc is significantly larger (or
smaller) than 50\%.
\paragraph{Not surprisingly, it helps to ask people to share.} (See Table
\ref{tb:test_call_action}; the notation for all tables is explained in
Table \ref{tb:notation}.) The basic sanity check we performed here was to take as
{\metric}s the number of occurrences of the verbs `rt', `retweet', `please', `spread', `pls',
and `plz' to capture explicit requests (e.g. ``please retweet'').  

\omt{
\begin{table}[t]
\centering
\caption{Explicit requests for sharing (where only occurrences
  POS-tagged as verbs count,
  according to the \protect\newcite{Gimpel+etal:11} tagger).
\label{tb:test_call_action}
}
  \ttablesize
  \begin{tabular}{|l|l|l|}
  \hline
& \eff & \su \\\hline
rt & $\uparrow\uparrow\uparrow\uparrow\white{}$~* &------\\
retweet & $\uparrow\uparrow\uparrow\uparrow\white{}$~* &YES~(59\%)\\
spread & $\uparrow\uparrow\uparrow\white{\uparrow}$~* &YES~(56\%)\\
please & $\uparrow\uparrow\uparrow\white{\uparrow}$~* &------\\
pls & $\uparrow\white{\uparrow\uparrow\uparrow}$ &------\\
plz & $\uparrow\uparrow\white{\uparrow\uparrow}$ &------\\

\hline
\end{tabular}
\end{table}
}

\paragraph{Informativeness helps.} (Table \ref{tb:test_informativeness}) Messages that are more informative have
increased {\em social exchange value} \cite{Homans:AmericanJournalOfSociology:1958},
and so may be
more worth propagating.  One crude approximation of
informativeness
is length,
and we see that length helps.\footnote{Of course,
simply inserting garbage
isn't going to lead to more retweets, but adding more information generally
involves longer text.}
In contrast, \newcite{simmons2011memes} found
that shorter versions
of memes
are more likely to be popular.
The difference may result from \tuc-pair changes being
more drastic than the variations that memes undergo.
\omt{
\begin{table}[t]
  \centering
  \caption{%
Informativeness.
  \label{tb:test_informativeness}}
  \ttablesize
  \begin{tabular}{|l|l|l|}
  \hline
  & \eff & \su \\\hline
length (chars) & $\uparrow\uparrow\uparrow\uparrow\white{}$~* &YES~(54\%)\\
verb & $\uparrow\uparrow\uparrow\uparrow\white{}$~* &YES~(56\%)\\
noun & $\uparrow\uparrow\uparrow\uparrow\white{}$~* &------\\
adjective & $\uparrow\uparrow\uparrow\white{\uparrow}$~* &YES~(51\%)\\
adverb & $\uparrow\uparrow\uparrow\white{\uparrow}$~* &YES~(55\%)\\
proper noun & $\uparrow\uparrow\uparrow\white{\uparrow}$~* &NO\white{--}~(45\%)\\
number & $\uparrow\uparrow\uparrow\uparrow\white{}$~* &NO\white{--}~(48\%)\\
hashtag & $\uparrow\white{\uparrow\uparrow\uparrow}$ &------\\
@-mention & $\downarrow\downarrow\downarrow\white{\downarrow}$~* &YES~(53\%)\\

\hline
  \end{tabular}
\end{table}
}

 A more refined informativeness measure is counts of the parts of speech
that correspond to content.
 Our POS results,
gathered using a Twitter-specific tagger \cite{Gimpel+etal:11},
  echo
those of
\newcite{Ashok+Feng+Choi:13} who looked at predicting the success of
books.
The diminished effect of hashtag inclusion with respect to what has been
reported previously \cite{Suh+etal:10,petrovic2011rt} presumably stems from our
topic and author controls.
\paragraph{Be like the community, and  be true to yourself (in the words you pick, but not necessarily in how you combine them).} (Table \ref{tb:test_conformity})
Although distinctive messages
 may attract attention, messages that conform to expectations might be more easily
accepted and therefore shared.  Prior work has explored this tension: \newcite{Lakkaraju+McAuley+Leskovec:13},
in a content-controlled study, found that the more up-voted
Reddit image titles balance novelty and familiarity;  \newcite{Danescu-Niculescu-Mizil+Cheng+Kleinberg+Lee:12} (henceforth
DCKL'12) showed that the memorability of movie quotes corresponds to higher
lexical distinctiveness but lower POS distinctiveness; and \newcite{Sun+Zhang+Mei:13} observed that deviating from one's own past language patterns correlates with more retweets.

Keeping in mind  that the authors in our data have at least 5000
followers\footnote{%
This is not an artificial restriction on our set of authors; a large follower count means (in
principle) that our results draw on a large sample of decisions whether to
retweet or not.},
we consider two types of language-conformity constraints an author might try to satisfy:
to be similar to what is normal
in the
Twitter community, and to be similar to what his or her followers expect.
We measure a tweet's similarity to expectations by its score
according to the relevant language model,
$\frac{1}{|T|}\sum_{x \in T}\log(p(x))$, where $T$ refers to
either
all the
unigrams (unigram model) or all and only bigrams (bigram model).\footnote{%
The tokens [at], [hashtag], [url] were ignored in the unigram-model case to prevent their
undue influence, but retained in the bigram model to capture longer-range usage
(``combination'') patterns.}
We trained a Twitter-community language model from
our 558M unpaired tweets, and personal language models
from each author's tweet history.

\begin{table}[t]
  \centering
  \caption{Conformity to the community and one's own past, measured via scores assigned by various language models.
  \label{tb:test_conformity}
  }
  \ttablesize
  \begin{tabular}{|l|l|l|}
  \hline
& \eff & \su \\\hline
twitter unigram & $\uparrow\uparrow\uparrow\white{\uparrow}$~* & YES~(54\%)\\
twitter bigram & $\uparrow\uparrow\uparrow\white{\uparrow}$~* & YES~(52\%)\\
  \hline
personal unigram & $\uparrow\uparrow\uparrow\white{\uparrow}$~* &YES~(52\%)\\
personal bigram & -------- &NO\white{--}~(48\%)\\
  \hline
  \end{tabular}
\end{table}

\paragraph{Imitate headlines.}
(Table \ref{tb:test_headline_model})
News headlines are
often
intentionally written
to be both informative and
attention-getting,
so we introduce the idea of scoring by a
language model built from New York Times headlines.\footnote {
To test whether the results stem from similarity to {\em news} rather than
headlines per se, we constructed a NYT-text LM, which proved less
effective.
We also
tried
using Gawker headlines (often
said to be
attention-getting)
but
pilot studies
revealed insufficient vocabulary overlap with our \tuc
pairs.
}

\omt{
\begin{table}[t]
  \centering
  \caption{
  LM-based resemblance to headlines.
  \label{tb:test_headline_model}}
  \ttablesize
  \begin{tabular}{|l|l|l|}
  \hline
 & \eff & \su \\\hline
headline unigram & $\uparrow\uparrow\white{\uparrow\uparrow}$ &YES~(53\%)\\
headline bigram & $\uparrow\uparrow\uparrow\uparrow\white{}$~* &YES~(52\%)\\

\hline
  \end{tabular}
\end{table}
}

\paragraph{Use words
associated with (non-paired) retweeted tweets.
} (Table \ref{tb:test_retweet_score})
We expect that provocative
or sensationalistic
tweets are likely to make people react.
We found it difficult to model provocativeness directly.  As a rough
approximation, we check whether the changes in $t_2$ with respect to $t_1$ (which
share the same
topic and author)
involve words or parts-of-speech that are associated with high retweet rate in
a very large separate sample of unpaired tweets (retweets and replies
discarded).  Specifically,
for each word $w$ that
appears more than 10 times, we compute the probability that tweets containing
$w$ are retweeted more than once%
, denoted
by $\rs(w)$.
We define the {\em \shortrs} of a tweet as $max_{w \in T} \rs(w)$,
where $T$ is all the words in the tweet,
and
the {\em \shortrs} of a
particular POS tag
$z$ in
a tweet as $max_{w \in T \& {\rm tag}(w)=z} \rs(w)$.

\omt{
\begin{table}[t]
  \centering
  \caption{Retweet score.
  \label{tb:test_retweet_score}}
  \ttablesize
  \begin{tabular}{|l|l|l|}
  \hline
  & \eff & \su \\\hline
\shortrs & $\uparrow\uparrow\white{\uparrow\uparrow}$~* &NO\white{--}~(49\%)\\
verb \shortrs & $\uparrow\uparrow\uparrow\uparrow\white{}$~* &------\\
noun \shortrs & $\uparrow\uparrow\uparrow\white{\uparrow}$~* &------\\
adjective \shortrs & $\uparrow\white{\uparrow\uparrow\uparrow}$ &YES~(50\%)\\
adverb \shortrs & $\uparrow\white{\uparrow\uparrow\uparrow}$ &YES~(51\%)\\
proper noun \shortrs & -------- &NO\white{--}~(48\%)\\

\hline
  \end{tabular}
\end{table}
}

\paragraph{Include positive and/or negative words.}
(Table \ref{tb:test_sentiment})
Prior work has found that including positive or negative sentiment increases message propagation
\cite{Berger+Milkman:12,Godes:MarketingLetters:2005,heath2001emotional,hansen2011good}.
We measured the occurrence of positive and negative words as
determined by the connotation lexicon of \newcite{Feng+Kang+Kuznetsova+Choi:13}
(better coverage than LIWC). Measuring the occurrence of both {\em
simultaneously} was inspired by
\newcite{Riloff+Qadir+Surve+DeSilva+Gilbert+Huang:13}.

\omt{
\begin{table}[t]
  \centering
  \caption{Sentiment (contrast is measured by presence of both
    positive and negative sentiments).
  \label{tb:test_sentiment}}
  \ttablesize
  \begin{tabular}{|l|l|l|}
  \hline
  & \eff & \su \\\hline
 positive & $\uparrow\uparrow\uparrow\white{\uparrow}$~* &------\\
 negative & $\uparrow\uparrow\uparrow\white{\uparrow}$~* &------\\
 contrast & $\uparrow\uparrow\uparrow\white{\uparrow}$~* &------\\

  \hline
  \end{tabular}
\end{table}
}

\paragraph{Refer to other people (but not your audience).}
(Table \ref{tb:test_pronoun})
First-person has been found useful for success before, but in the
different domains of scientific abstracts
\cite{Guerini:ProceedingsOfIcwsm:2012} and books
\cite{Ashok+Feng+Choi:13}.
\omt{
\begin{table}[t]
  \centering
  \caption{%
Pronouns.
  \label{tb:test_pronoun}}
 \ttablesize
  \begin{tabular}{|l|l|l|}
  \hline
  & \eff & \su \\\hline\hline
1st person singular & -------- &YES~(51\%)\\
1st person plural & -------- &YES~(52\%)\\
2nd person & -------- &YES~(57\%)\\
3rd person singular & $\uparrow\uparrow\white{\uparrow\uparrow}$ &YES~(55\%)\\
3rd person plural & $\uparrow\white{\uparrow\uparrow\uparrow}$ &YES~(58\%)\\

\hline
 \end{tabular}
\end{table}
}

\paragraph{Generality helps.}
(Table \ref{tb:test_generality})
DCKL'12 posited that movie quotes are more shared in the culture when they are general
enough to be used in multiple contexts.  We hence measured the presence
of indefinite articles vs. definite articles.

\omt{
\begin{table}[t]
  \centering
  \caption{Generality.\label{tb:test_generality}}
 \ttablesize
  \begin{tabular}{|l|l|l|}
  \hline
& \eff & \su \\\hline
indefinite articles (a,an) & $\uparrow\uparrow\uparrow\white{\uparrow}$~* &------\\
definite articles (the) & -------- &YES~(52\%)\\

\hline
\end{tabular}
\end{table}
}

\paragraph{The easier to read, the better.}
(Table \ref{tb:test_readability})
We measure readability by using Flesch reading ease
\cite{flesch1948new} and
Flesch-Kincaid  grade level \cite{kincaid1975derivation},
though 
they are not designed for short texts.
We use negative grade level so that a larger
value indicates
easier texts to read.
 \omt{
\begin{table}[t]
  \centering
  \caption{Readability.
 \label{tb:test_readability}
  }
 \ttablesize
\begin{tabular}{|l|l|l|}
\hline
& \eff & \su \\\hline
reading ease & $\uparrow\uparrow\white{\uparrow\uparrow}$ &YES~(52\%)\\
negative grade level & $\uparrow\white{\uparrow\uparrow\uparrow}$ &YES~(52\%)\\

\hline
\end{tabular}
\end{table}
 }

\vspace{-0.2in}
\paragraph{Final question: Do authors prefer to do what is effective?}
Recall that we use binomial tests to determine author preference for
applying a feature more in $t_2$.
Our preference statistics show that author
preferences in many cases are aligned with feature efficacy.
But there are several notable exceptions: for example, authors tend to increase the use of @-mentions
and 2nd person pronouns even though they are ineffective.
On the other hand, they did not
increase the use of
effective ones like proper nouns and
numbers; nor did they tend to increase their rate of
sentiment-bearing words.
Bearing in mind that changes in $t_2$ may not always be intended as an effort to improve $t_1$, it is
still interesting to observe that there are some contrasts between feature efficacy and
author preferences.

\newcommand{\bigram}{1,2-gram\xspace}
\newcommand{\baseline}{{$\neg$\tuc}+ff+time\xspace}%
\newcommand{\Baseline}{\baseline}

\subsection{Predicting the ``better'' wording}
\label{sec:pred}

\begin{figure*}[t]
\centering
\subfigure[
Cross-validation and heldout accuracy
  for  various feature sets.
  Blue lines inside bars: performance
  when custom features are restricted to those that pass our Bonferroni correction (no line for readability because no readability features passed).
  Dashed vertical line: \baseline performance.
]{
  \includegraphics[width=0.64\textwidth]{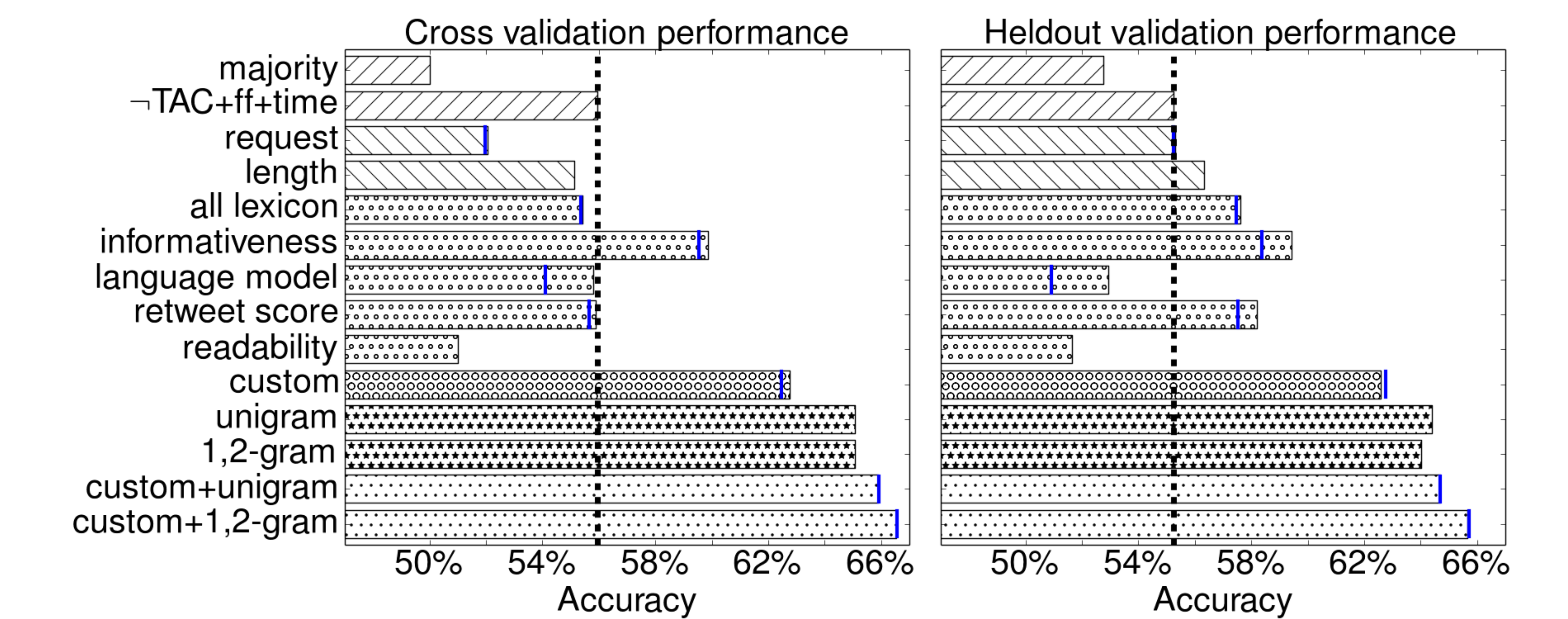}
  \label{fig:performance_comparison}
}
\hfill
\subfigure[
  Cross-validation
  accuracy  vs data size.
  Human performance  was estimated from a disjoint set of
  100 pairs (see \S\ref{sec:turkers}).
  ]{
  \includegraphics[width=0.32\textwidth]{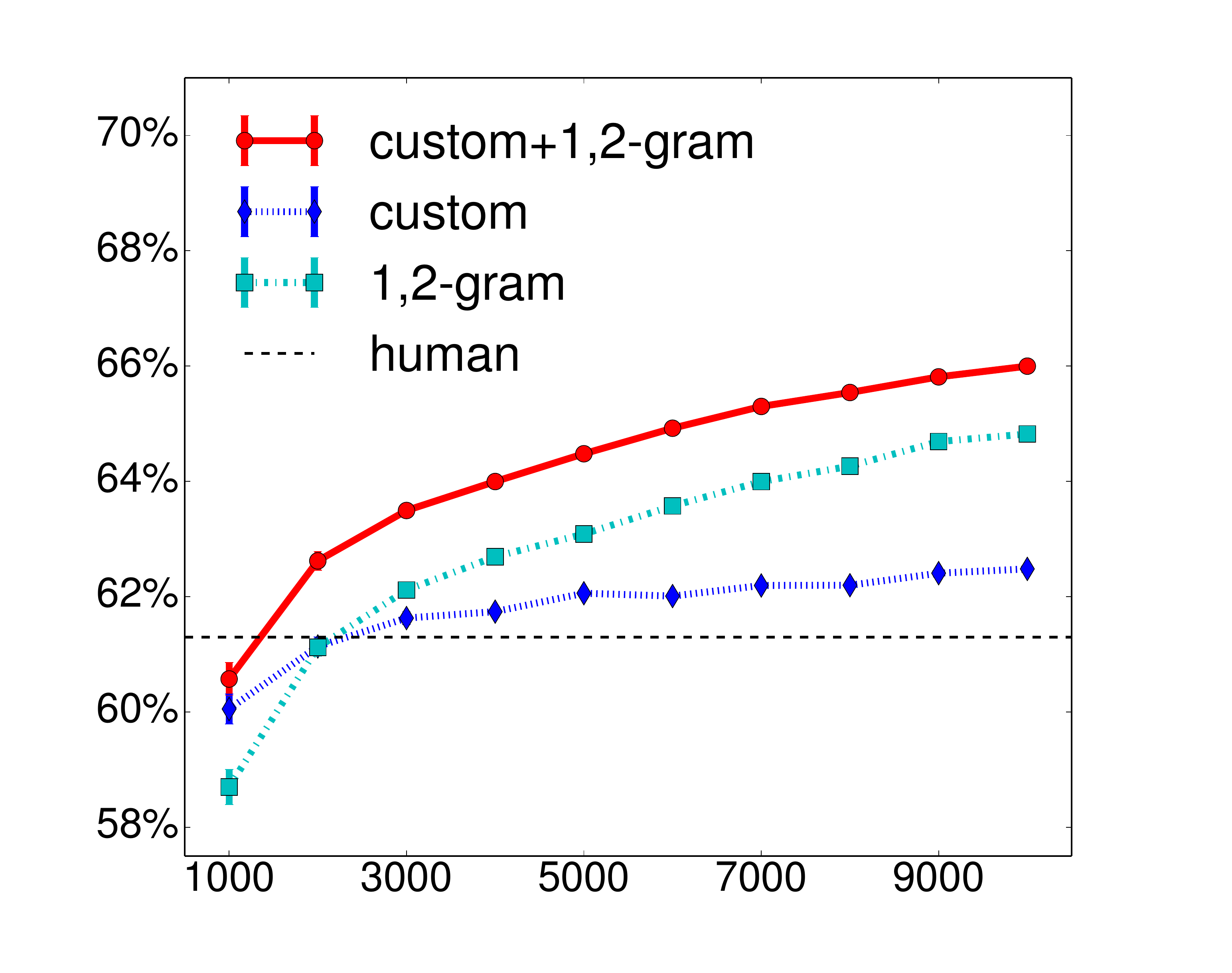}
  \label{fig:performance_size}
}
\caption{
  Accuracy results.
  Pertinent significance results are as follows.
  In cross-validation, custom+\bigram is significantly better than \baseline (p=0) and \bigram (p=3.8e-7).
  In heldout validation,
  custom+\bigram is significantly better than \baseline (p=3.4e-12) and \bigram (p=0.01) but not
  unigram (p=0.08),
  perhaps due to the small size of the heldout set.
\label{fig:performance}}
\end{figure*}

Here,
we
further
examine the collective efficacy of the features introduced in \S \ref{sec:testing}
via their performance on a binary prediction task: given a \tuc pair ($t_1, t_2$), did $t_2$ receive
more retweets?

\paragraph{Our approach.}
We group the features introduced in \S \ref{sec:testing} into 16
lexicon-based features (Table \ref{tb:test_call_action}, \ref{tb:test_sentiment}, \ref{tb:test_pronoun}, \ref{tb:test_generality}), 9
informativeness features (Table \ref{tb:test_informativeness}),
6 language model
features (Table \ref{tb:test_conformity},
\ref{tb:test_headline_model}),
6 \shortrs features (Table \ref{tb:test_retweet_score}), and
2 readability features (Table \ref{tb:test_readability}). We refer to all 39 of them
together
 as
{\em custom} features.  We also consider
tagged
bag-of-words (``BOW'') features,
which includes all the unigram (word:POS pair) and bigram features that appear more than 10 times in %
the cross-validation data.
This yields 3,568 unigram features and
4,095 bigram features, for a total of 7,663
so-called
{\em \mbox{\bigram} features}.
Values for each feature are normalized
by linear transformation
across all tweets in the training data to lie in the range $[0,1]$.\footnote{%
  We also tried normalization by {\em whitening},
  but it did
  not lead to further improvements.}

For a given \tuc pair, we construct its feature vector as follows.
For each feature being considered, we compute its normalized value for each tweet in the pair
and take the difference as the feature value for this pair.
We use L2-regularized logistic regression
as
our classifier,
with parameters chosen by cross validation on
the training data.
(We also experimented with SVMs. The performance was very close, but
  mostly slightly lower.)
\paragraph{A strong non-\tuc alternative, with social information and timing thrown in.}
One baseline result we would like to establish is whether the topic and author controls we have argued
for, while intuitively compelling for the purposes of trying to determine the best way for a given author to present some fixed content, are really necessary in practice.
To test this, we consider an alternative binary L2-regularized logistic-regression classifier that is trained on
unpaired data, specifically, on the collection of 10,000 most retweeted tweets
(gold-standard label: positive)
plus the 10,000 least retweeted tweets (gold-standard label:
negative)
that are neither retweets nor replies%
.
Note
that this alternative thus is granted, by design, roughly {\em twice} the training instances
that our classifiers have, as a result of having roughly the same number of
tweets, since our instances are pairs.  Moreover, we additionally include the tweet
author's follower count, and the day and hour of posting, as features.
We refer to this alternative classifier as \mbox{{\em \baseline}}. (Mnemonic: ``ff'' is
used
in
bibliographic contexts
as an abbreviation for ``and the following''.)  We apply it
to a tweet pair by computing whether it gives a higher score to $t_2$ or not.

\paragraph{Baselines.}
To sanity-check whether our classifier provides any improvement over the simplest
methods one could try, we also
report the performance of the majority
baseline,
our request-for-sharing features, and our character-length feature.

\paragraph{Performance comparison.}
We compare the accuracy (percentage of pairs
whose labels were correctly predicted)
of our approach
against the
competing
methods.
We report 5-fold cross validation results on our
balanced set of
11,404 \tuc pairs
and on our completely disjoint heldout data\footnote{%
To construct this data, we used the same criteria as in \S\ref{sec:data}:
written by authors with more than 5000 followers, posted within 12 hours, $n_2-n_1\geq 10$ or $\leq -15$, and
cosine similarity
threshold value the same as in \S\ref{sec:data},
cap of 50 on number of pairs from any individual author.}
of 1,770 \tuc pairs;
this set was never examined during development, and there are no authors in
common between the two testing sets.

Figure \ref{fig:performance_comparison} summarizes the main results.
While \baseline outperforms the
majority baseline,
using all the features we proposed
beats \baseline by
more than 10\% in both cross-validation (66.5\% vs 55.9\%) and heldout
validation (65.6\% vs 55.3\%).
We outperform the average human accuracy of 61\%
reported in our Amazon Mechanical Turk experiments (for a different data sample); \baseline fails to do so.

The importance of topic and author control can be seen by further investigation
of \baseline's performance. First,
note that
it
yields an accuracy of
around
55\%
on our
alternate-version-selection
task,\footnote{%
One might suspect that the problem is that \baseline learns from its training data to over-rely on follower-count, since that is presumably a good feature for non-\tuc tweets, and for this reason suffers when run on \tuc data where follower-counts are by construction non-informative.  But in fact, we found that removing the follower-count feature from \baseline and re-training did not lead to improved performance.  Hence, it seems that it is the  non-controlled nature of the alternate training data that explains the drop in performance.
}
even though its
cross-validation accuracy
on the larger
most- and least-retweeted
unpaired
 tweets averages
out
to a high 98.8\%.
Furthermore, note the superior performance of
unigrams
trained on \tuc data vs
{\baseline} --- which is similar to our unigrams
but trained on a
larger but non-\tuc  dataset that included
metadata. %
Thus, \tuc pairs are a useful data source even for non-custom features.
(We also include
individual feature comparisons later.)
Informativeness is the
best-performing
custom
 feature group
when run in isolation,
 and outperforms all baselines, as well as \baseline;
 and we can see from Figure \ref{fig:performance_comparison} that this is not
 due just to length.
The combination of all our 39 custom features yields approximately 63\% accuracy in
both testing settings, significantly outperforming informativeness alone (p$<$0.001 in both
cases). Again, this is
higher than
our estimate of average human performance.

Not surprisingly,
the \tuc-trained
BOW features
(unigram and \bigram)
show impressive predictive power in this
task: many of our custom features
can be %
captured by
bag-of-word features, in a way.
Still, the best performance is achieved by combining our custom and \bigram features together,
to a degree
statistically significantly better than using
\bigram features alone.
Finally, we remark on our Bonferroni correction.
 Recall that the intent
of applying
it
is to avoid false positives. However, in our case,
Figure \ref{fig:performance_comparison} shows that our potentially ``false'' positives --- features
whose effectiveness did not pass the Bonferroni correction test --- actually do raise performance in our prediction tests.

\paragraph{Size of training data.}
Another interesting observation is how performance varies with data size.
For $n=1000, 2000, \ldots, 10000$, we
randomly sampled $n$ pairs from our 11,404 pairs, and computed
the average cross-validation accuracy
on the sampled data.
Figure \ref{fig:performance_size} shows the averages over 50 runs of the
aforementioned procedure.
Our custom features can achieve good performance with little data, in the sense
that for sample size $1000$, they outperform BOW features; on the other hand,
BOW features quickly surpass them.
Across the board,
the custom+\bigram features are
consistently better than the \bigram
features
alone.

\paragraph{Top features.} Finally, we examine
some of
the
top-weighted individual
features
from our approach and from the
competing
\baseline
classifier.
The top three rows of
Table \ref{tb:word_coefficients} show the best \custom and best and worst
unigram
features
for our method;
the bottom two rows show the best and worst unigrams for \baseline.
Among \custom features,
we see that community and personal language models, informativeness, retweet scores, sentiment, and
generality are represented.
As for unigram features, not surprisingly, ``rt''
and
 ``retweet'' are top
features
for both our approach and \baseline.
However, the other unigrams for the two methods seem to be a bit different in
spirit.  Some of the unigrams determined to be most poor only by our method
appear to be both surprising and yet plausible in retrospect: ``icymi'' (abbreviation for ``in case
you missed it'') tends to indicate a direct repetition of older information, so people
might prefer to retweet the earlier version; ``thanks'' and ``sorry'' could
correspond to personal thank-yous and apologies not meant to be shared with a
broader audience, and similarly @-mentioning another user may indicate a tweet intended only for that person.
The appearance of [hashtag] in the best \baseline unigrams is consistent with
prior research in non-\tuc settings \cite{Suh+etal:10,petrovic2011rt}.

\begin{table}[t]
\centering
\caption{
Features with largest coefficients, delimited by commas.  POS tags omitted for clarity.
\label{tb:word_coefficients}}
\vspace{-0.1in}
\ttablesize
\begin{tabular}{|p{7cm}|}
\hline
\multicolumn{1}{|c|}{Our approach}\\
\hline
{\bf best 15 custom} \hspace{0.5cm} twitter bigram, length (chars), rt (the word), retweet (the word), verb, verb retweet score, personal unigram, proper noun, number, noun, positive words, please (the word), proper noun retweet score, indefinite articles (a,an), adjective\\
{\bf best 20 unigrams} \hspace{0.5cm} rt, retweet, [num], breaking, is, win, never, ., people, need, official, officially, are, please, november, world, girl, !!!, god, new\\
{\bf worst 20 unigrams} \hspace{0.5cm} :, [at], icymi, also, comments, half, ?, earlier, thanks, sorry, highlights, bit, point, update, last, helping, peek, what, haven't, debate\\

\hline\hline
\multicolumn{1}{|c|}{
\Baseline}\\
\hline
{\bf best 20 unigrams} \hspace{0.5cm} [hashtag], teen, fans, retweet, sale, usa, women, butt, caught, visit, background, upcoming, rt, this, bieber, these, each, chat, houston, book\\
{\bf worst 20 unigrams} \hspace{0.5cm} :, ..., boss, foundation, ?, $\sim$, others, john, roll, ride, appreciate, page, drive, correct, full, ', looks, @ (not as [at]), sales, hurts\\

\hline
\end{tabular}
\end{table}

\section{Conclusion}
\label{sec:conclusion}

In this work, we conducted the first
large-scale topic- and author-controlled
experiment
to study the effects of wording on information propagation.
The features we developed to choose the better of two alternative wordings posted better performance than that of all our comparison algorithms, including one given access to author and timing features but trained on non-\tuc data, and also bested our estimate of average human performance.
According to our hypothesis tests,
helpful wording heuristics include
adding more information,
making one's language align with both community norms and with one's prior messages,
and mimicking news headlines.
Readers may try out their own alternate phrasings at
\url{http://chenhaot.com/retweetedmore/} to see
what a simplified version of our classifier predicts.

In future work, it will be
interesting to examine how these features generalize to longer and
more extensive arguments.
Moreover, understanding the underlying psychological and cultural mechanisms that
establish the effectiveness of these features is a fundamental problem
of interest.

\newcommand{\finit}[2]{#1.}
\noindent{\bf Acknowledgments.} We thank
\finit{C}{hris} Callison-Burch,
\finit{C}{ristian} Danescu-Niculescu-Mizil,
\finit{J}{on} Kleinberg,
\finit{P}{arvaz} Mahdabi,
\finit{S}{endhil} Mullainathan,
\finit{F}{ernando} Pereira,
\finit{K}{arthik} Raman,
\finit{A}{dith} Swaminathan,
the Cornell NLP seminar participants and the
reviewers for their comments;
\finit{J}{ure} Leskovec for providing some initial data;
and the anonymous annotators for all their labeling help.
This work was supported in part by  NSF grant IIS-0910664 and a Google Research Grant.
\normalsize

\bibliographystyle{acl}
\bibliography{small}

\begin{thebibliography}{}

\bibitem[\protect\citename{Artzi \bgroup et al.\egroup }2012]{artzipredicting}
Yoav Artzi, Patrick Pantel, and Michael Gamon.
\newblock 2012.
\newblock Predicting responses to microblog posts.
\newblock In {\em Proceedings of NAACL (short paper)}.

\bibitem[\protect\citename{Ashok \bgroup et al.\egroup
  }2013]{Ashok+Feng+Choi:13}
Vikas~Ganjigunte Ashok, Song Feng, and Yejin Choi.
\newblock 2013.
\newblock {Success with style: Using writing style to predict the success of
  novels}.
\newblock In {\em Proceedings of EMNLP}.

\bibitem[\protect\citename{Bakshy \bgroup et al.\egroup
  }2011]{Bakshy:ProceedingsOfWsdm:2011}
Eitan Bakshy, Jake~M. Hofman, Winter~A. Mason, and Duncan~J. Watts.
\newblock 2011.
\newblock Everyone's an influencer: Quantifying influence on twitter.
\newblock In {\em Proceedings of WSDM}.

\bibitem[\protect\citename{Bell and Sejnowski}1997]{bell1997independent}
Anthony~J. Bell and Terrence~J. Sejnowski.
\newblock 1997.
\newblock {The independent components of natural scenes are edge filters}.
\newblock {\em Vision research}, 37(23):3327--3338.

\bibitem[\protect\citename{Benjamini and
  Hochberg}1995]{benjamini1995controlling}
Yoav Benjamini and Yosef Hochberg.
\newblock 1995.
\newblock {Controlling the false discovery rate: {A} practical and powerful
  approach to multiple testing}.
\newblock {\em Journal of the Royal Statistical Society. Series B
  (Methodological)}, pages 289--300.

\bibitem[\protect\citename{Borghol \bgroup et al.\egroup
  }2012]{Borghol+etal:12}
Youmna Borghol, Sebastien Ardon, Niklas Carlsson, Derek Eager, and Anirban
  Mahanti.
\newblock 2012.
\newblock The untold story of the clones: Content-agnostic factors that impact
  {YouTube} video popularity.
\newblock In {\em Proceedings of KDD}.

\bibitem[\protect\citename{Chong and Druckman}2007]{Chong+Druckman:2007a}
Dennis Chong and James~N. Druckman.
\newblock 2007.
\newblock {Framing theory}.
\newblock {\em Annual Review of Political Science}, 10:103--126.

\bibitem[\protect\citename{Danescu-Niculescu-Mizil \bgroup et al.\egroup
  }2012]{Danescu-Niculescu-Mizil+Cheng+Kleinberg+Lee:12}
Cristian Danescu-Niculescu-Mizil, Justin Cheng, Jon Kleinberg, and Lillian Lee.
\newblock 2012.
\newblock {You had me at hello: {How} phrasing affects memorability}.
\newblock In {\em Proceedings of ACL}.

\bibitem[\protect\citename{DiNardo}2008]{dinardo2008natural}
John DiNardo.
\newblock 2008.
\newblock Natural experiments and quasi-natural experiments.
\newblock In {\em The New Palgrave Dictionary of Economics}. Palgrave
  Macmillan.

\bibitem[\protect\citename{Dunn}1961]{dunn1961multiple}
Olive~Jean Dunn.
\newblock 1961.
\newblock {Multiple comparisons among means}.
\newblock {\em Journal of the American Statistical Association},
  56(293):52--64.

\bibitem[\protect\citename{Feng \bgroup et al.\egroup
  }2013]{Feng+Kang+Kuznetsova+Choi:13}
Song Feng, Jun~Seok Kang, Polina Kuznetsova, and Yejin Choi.
\newblock 2013.
\newblock {Connotation lexicon: A dash of sentiment beneath the surface
  meaning}.
\newblock In {\em Proceedings of ACL}.

\bibitem[\protect\citename{Flesch}1948]{flesch1948new}
Rudolph Flesch.
\newblock 1948.
\newblock {A new readability yardstick}.
\newblock {\em Journal of applied psychology}, 32(3):221.

\bibitem[\protect\citename{Gimpel \bgroup et al.\egroup }2011]{Gimpel+etal:11}
Kevin Gimpel, Nathan Schneider, Brendan O'Connor, Dipanjan Das, Daniel Mills,
  Jacob Eisenstein, Michael Heilman, Dani Yogatama, Jeffrey Flanigan, and
  Noah~A. Smith.
\newblock 2011.
\newblock {Part-of-speech Tagging for Twitter: Annotation, Features, and
  Experiments}.
\newblock In {\em Proceedings of NAACL (short paper)}.

\bibitem[\protect\citename{Godes \bgroup et al.\egroup
  }2005]{Godes:MarketingLetters:2005}
David Godes, Dina Mayzlin, Yubo Chen, Sanjiv Das, Chrysanthos Dellarocas, Bruce
  Pfeiffer, Barak Libai, Subrata Sen, Mengze Shi, and Peeter Verlegh.
\newblock 2005.
\newblock {The firm's management of social interactions}.
\newblock {\em Marketing Letters}, 16(3-4):415--428.

\bibitem[\protect\citename{Guerini \bgroup et al.\egroup
  }2011]{Guerini:ProceedingsOfIcwsm:2011}
Marco Guerini, Carlo Strapparava, and G\"ozde \"Ozbal.
\newblock 2011.
\newblock Exploring text virality in social networks.
\newblock In {\em Proceedings of ICWSM (poster)}.

\bibitem[\protect\citename{Guerini \bgroup et al.\egroup
  }2012]{Guerini:ProceedingsOfIcwsm:2012}
Marco Guerini, Alberto Pepe, and Bruno Lepri.
\newblock 2012.
\newblock Do linguistic style and readability of scientific abstracts affect
  their virality?
\newblock In {\em Proceedings of ICWSM (poster)}.

\bibitem[\protect\citename{Hansen \bgroup et al.\egroup }2011]{hansen2011good}
Lars~Kai Hansen, Adam Arvidsson, Finn~\r{A}rup Nielsen, Elanor Colleoni, and
  Michael Etter.
\newblock 2011.
\newblock {Good friends, bad news-affect and virality in Twitter}.
\newblock {\em Communications in Computer and Information Science}, 185:34--43.

\bibitem[\protect\citename{Heath \bgroup et al.\egroup
  }2001]{heath2001emotional}
Chip Heath, Chris Bell, and Emily Sternberg.
\newblock 2001.
\newblock {Emotional selection in memes: The case of urban legends}.
\newblock {\em Journal of personality and social psychology}, 81(6):1028.

\bibitem[\protect\citename{Homans}1958]{Homans:AmericanJournalOfSociology:1958}
George~C. Homans.
\newblock 1958.
\newblock {Social Behavior as Exchange}.
\newblock {\em American Journal of Sociology}, 63(6):597--606.

\bibitem[\protect\citename{Hong \bgroup et al.\egroup
  }2011]{Hong:2011:PPM:1963192.1963222}
Liangjie Hong, Ovidiu Dan, and Brian~D. Davison.
\newblock 2011.
\newblock Predicting popular messages in {Twitter}.
\newblock In {\em Proceedings of WWW}.

\bibitem[\protect\citename{Hovland \bgroup et al.\egroup
  }1953]{Hovland+Janis+Kelley:54}
Carl~I. Hovland, Irving~L. Janis, and Harold~H. Kelley.
\newblock 1953.
\newblock {\em Communication and Persuasion: Psychological Studies of Opinion
  Change}, volume~19.
\newblock Yale University Press.

\bibitem[\protect\citename{Kincaid \bgroup et al.\egroup
  }1975]{kincaid1975derivation}
J.~Peter Kincaid, Robert~P. Fishburne~Jr., Richard~L. Rogers, and Brad~S.
  Chissom.
\newblock 1975.
\newblock Derivation of new readability formulas (automated readability index,
  fog count and flesch reading ease formula) for navy enlisted personnel.
\newblock Technical report, DTIC Document.

\bibitem[\protect\citename{Kwak \bgroup et al.\egroup
  }2010]{Kwak+Lee+Park+Moon:10}
Haewoon Kwak, Changhyun Lee, Hosung Park, and Sue Moon.
\newblock 2010.
\newblock {What is Twitter, a social network or a news media?}
\newblock In {\em Proceedings of WWW}.

\bibitem[\protect\citename{Lakkaraju \bgroup et al.\egroup
  }2013]{Lakkaraju+McAuley+Leskovec:13}
Himabindu Lakkaraju, Julian McAuley, and Jure Leskovec.
\newblock 2013.
\newblock {What's in a name? Understanding the interplay between titles,
  content, and communities in social media}.
\newblock In {\em Proceedings of ICWSM}.

\bibitem[\protect\citename{Louis and Nenkova}2013]{Louis+Nenkova:13}
Annie Louis and Ani Nenkova.
\newblock 2013.
\newblock {What makes writing great? First experiments on article quality
  prediction in the science journalism domain}.
\newblock {\em Transactions of ACL}.

\bibitem[\protect\citename{Ma \bgroup et al.\egroup }2012]{ma2012will}
Zongyang Ma, Aixin Sun, and Gao Cong.
\newblock 2012.
\newblock {Will this \#hashtag be popular tomorrow?}
\newblock In {\em Proceedings of SIGIR}.

\bibitem[\protect\citename{McIntyre and
  Lapata}2009]{McIntyre:2009:LTT:1687878.1687910}
Neil McIntyre and Mirella Lapata.
\newblock 2009.
\newblock {Learning to tell tales: A data-driven approach to story generation}.
\newblock In {\em Proceedings of ACL-IJCNLP}.

\bibitem[\protect\citename{Milkman and Berger}2012]{Berger+Milkman:12}
Katherine~L Milkman and Jonah Berger.
\newblock 2012.
\newblock {What makes online content viral?}
\newblock {\em Journal of Marketing Research}, 49(2):192--205.

\bibitem[\protect\citename{Petrovi\'{c} \bgroup et al.\egroup
  }2011]{petrovic2011rt}
Sa\v{s}a Petrovi\'{c}, Miles Osborne, and Victor Lavrenko.
\newblock 2011.
\newblock {RT} to win! {Predicting} message propagation in {Twitter}.
\newblock In {\em Proceedings of ICWSM}.

\bibitem[\protect\citename{Pitler and Nenkova}2008]{Pitler:2008}
Emily Pitler and Ani Nenkova.
\newblock 2008.
\newblock {Revisiting readability: A unified framework for predicting text
  quality}.
\newblock In {\em Proceedings of EMNLP}.

\bibitem[\protect\citename{Riloff \bgroup et al.\egroup
  }2013]{Riloff+Qadir+Surve+DeSilva+Gilbert+Huang:13}
Ellen Riloff, Ashequl Qadir, Prafulla Surve, Lalindra De~Silva, Nathan Gilbert,
  and Ruihong Huang.
\newblock 2013.
\newblock {Sarcasm as contrast between a positive sentiment and negative
  situation}.
\newblock In {\em Proceedings of EMNLP}.

\bibitem[\protect\citename{Romero \bgroup et al.\egroup
  }2013]{Romero+Tan+Ugander:13}
Daniel~M. Romero, Chenhao Tan, and Johan Ugander.
\newblock 2013.
\newblock {On the interplay between social and topical structure}.
\newblock In {\em Proceedings of ICWSM}.

\bibitem[\protect\citename{Salganik \bgroup et al.\egroup
  }2006]{Salganik:Science:2006}
Matthew~J. Salganik, Peter~Sheridan Dodds, and Duncan~J. Watts.
\newblock 2006.
\newblock Experimental study of inequality and unpredictability in an
  artificial cultural market.
\newblock {\em Science}, 311(5762):854--856.

\bibitem[\protect\citename{Simmons \bgroup et al.\egroup
  }2011]{simmons2011memes}
Matthew~P. Simmons, Lada~A Adamic, and Eytan Adar.
\newblock 2011.
\newblock {Memes online: Extracted, subtracted, injected, and recollected}.
\newblock In {\em Proceedings of ICWSM}.

\bibitem[\protect\citename{Suh \bgroup et al.\egroup }2010]{Suh+etal:10}
Bongwon Suh, Lichan Hong, Peter Pirolli, and Ed~H. Chi.
\newblock 2010.
\newblock {Want to be retweeted? {Large} scale analytics on factors impacting
  retweet in {Twitter} network}.
\newblock In {\em Proceedings of SocialCom}.

\bibitem[\protect\citename{Sun \bgroup et al.\egroup }2013]{Sun+Zhang+Mei:13}
Tao Sun, Ming Zhang, and Qiaozhu Mei.
\newblock 2013.
\newblock {Unexpected relevance: An empirical study of serendipity in
  retweets}.
\newblock In {\em Proceedings of ICWSM}.

\bibitem[\protect\citename{Tsur and Rappoport}2012]{Tsur+Rappoport:12}
Oren Tsur and Ari Rappoport.
\newblock 2012.
\newblock {What's in a hashtag?: Content based prediction of the spread of
  ideas in microblogging communities}.
\newblock In {\em Proceedings of WSDM}.

\bibitem[\protect\citename{Yang and
  Leskovec}2011]{Yang:2011:PTV:1935826.1935863}
Jaewon Yang and Jure Leskovec.
\newblock 2011.
\newblock Patterns of temporal variation in online media.
\newblock In {\em Proceedings of WSDM}.

\end{thebibliography}

\end{document}